\documentclass[a4paper,10pt]{article}
\usepackage[utf8]{inputenc}

\title{  Consequences of  Deformation of the Heisenberg Algebra}
\author{Mir Faizal \\Department of Physics and Astronomy, \\  University of Waterloo,   Waterloo,\\
Ontario N2L 3G1, Canada 
}
\date{}
\begin{document}

\maketitle

\begin{abstract}
In this paper we will demonstrate that like the existence of a minimum measurable length, 
the existence of a maximum measurable momentum, also influence all quantum mechanical systems. 
Beyond the simple one dimensional case, the existence of 
a maximum momentum will induce  non-local  
corrections to the first quantized Hamiltonian. However, these non-local corrections 
can be effectively treated as 
local corrections by using the theory of  harmonic extensions of functions. 
We will also analyses the second quantization of this deformed
first quantized theory. Finally, we will analyses the gauge symmetry 
corresponding to this deformed theory.
\end{abstract}
\section{Introduction}
The classical picture of spacetime gets modified in all most all 
approaches to quantum gravity. One of the most common modification that 
 occurs in most theories of quantum gravity is that the 
the continuum picture of spacetime breaks down at Planck scale. 
So, the idea of the existence of a minimum length is naturally incorporated  
in most approaches to quantum gravity. A minimum measurable 
length naturally occurs in string theory \cite{z2}-\cite{2z}. Even in loop quantum gravity  
the existence of minimum length 
 turns big bang into a big bounce \cite{z1}. In fact, there are 
strong indication  from black hole physics that there 
minimum length of the order of the Planck length should 
arise in any theory of quantum gravity \cite{z4}-\cite{z5}.
The existence of a minimum measurable length deforms 
all quantum mechanical Hamiltonians  by 
$
 H \psi = H_0\psi + H_1\psi
$, where $H_0 =    p^2/2m + V (x) $ is the original Hamiltonian, $   p_i = -i \partial_i$, 
is the momentum operator corresponding to the undeformed Heisenberg algebra, and 
$H_1 = \beta    p^4 /m$ is the term that occurs due to the existence of a  minimum length.

According to the Heisenberg uncertainty principle there
is no limit to the accuracy with which one can measure 
 the momentum or the position of a particle separately. 
 In other words, according to the Heisenberg uncertainty principle
minimum observable length is actually zero. 
Thus, the Heisenberg uncertainty principle has to be modified in order to incorporate
the idea of minimum length. 
This has been done and the resultant uncertainty principle is called the 
 Generalized Uncertainty Principle 
\cite{1}-\cite{54}. In this picture  the commutation relations between position
and momentum operators in the Hilbert space are  deformed to contain momentum dependent factors.

A further deformation of this algebra occurs in doubly special relativity \cite{2}-\cite{3}. 
In doubly special special relativity theories, both the 
 velocity of light and the Planck energy are
invariant quantities. This modified algebra naturally incorporates the existence of a
maximal momentum
for any particle. In fact, General Relativity has also been modified to keep
both velocity of light and the Planck energy
as invariant quantities. The resultant theory is called Gravity's Rainbow \cite{n1}-\cite{n2}. 
Both these deformations have been combined into a single deformation of 
the Heisenberg algebra \cite{n4}-\cite{n5}. 
The modification to transition rate of ultra cold neutrons in gravitational field
has been studied in this deformed algebra \cite{n6}. 
In fact,  the modification to the Lamb shift and Landau levels have
also been analysed in this deformed algebra \cite{n7}. 
However, both these calculations are only done for the simple one dimensional case. 
In this paper we will give a method for analyzing these deformations for higher dimensional cases. 
We will also analyses the second quantization of this deformed first quantized theory. 
We will finally construct a gauge theory corresponding to 
this deformed quantum field theory. 
It will be observed that even though the Lagrangian for the 
gauge theories will be non-local, the gauge symmetry will  be generated 
by local gauge transformations.
Using this fact, these non-local gauge theories will 
be  quantized. 
\section{ Deformed Heisenberg Algebra}
  The deformation of the  Heisenberg 
algebra consistent with the existence of a minimum length is  $[x^i, p_j ] = 
i [\delta_{j}^i + \beta p^2 \delta_{j}^i + 2 \beta p^i p_j]$ and the 
deformation of the Heisenberg algebra consistent with the existence of a maximum momentum is 
$ [x^i, p_j] =
i [1 - \tilde \beta |p| \delta_{j}^i + \tilde \beta p^i p_j]$, with
$\beta = \beta_0 \ell_{Pl}/ \hbar$,
and $\tilde \beta = \ell_{Pl}$, where $\ell_{Pl}\approx 10^{-35}~m$ is the Planck length,
 and $\beta_0$ is a constant normally assumed to be of order unity. 
Both these deformations of Heisenberg algebra can be combined into a single 
deformation of the  Heisenberg algebra  as follows \cite{n4}-\cite{n5}
\begin{equation}
 [x^i, p_j] = i \left[  \delta_{j}^i - \alpha |p| \delta_{j}^i + \alpha |p|^{-1} p^i p_j 
  + \alpha^2 p^2 \delta_{j}^i + 3 \alpha^2 p^i p_j\right],
\end{equation}
with $\alpha = {\alpha_0}/{M_{Pl}c} = {\alpha_0 \ell_{Pl}}/{\hbar},$ where 
$M_{Pl}$ is the Planck mass, $\ell_{Pl}\approx 10^{-35}~m$ is the Planck length,
and $M_{Pl} c^2 \approx 10^{19}~GeV$ is the Planck energy. In the one dimensional case this 
corresponds to the uncertainty relation given by $\Delta x \Delta p = 
[1 - 2 \alpha <p> + 4 \alpha^2 <p^2> ]$. These imply the existence of a minimum length  
\begin{equation}
 \Delta x \geq \Delta x_{min} \geq \alpha_0 \ell_{Pl}.
\end{equation}
It also implies the existence of a maximum 
momentum 
\begin{equation}
 \Delta p \leq \Delta p_{max} \leq \alpha_0^{-1} M_{Pl} c.
\end{equation}
In fact, this algebra is satisfied if 
we chose the following representation for it, $ x_i = \tilde x_i $ and $p_i = \tilde p_i 
(1 - \alpha |\tilde p|  + 2\alpha^2 \tilde p^2)$, such that $[\tilde x^j, \tilde p_i] 
= i \delta^j_i$. Thus, $\tilde p_i$ can be interpreted as the momentum at low energies. 
Now can use the standard representation for $\tilde p$ in one dimension and  obtain 
the following expression for $p$, 
\begin{equation}
p = -i\left(1 + i \alpha \frac{d}{dx} - 2\alpha^2 \frac{d^2}{d^2x}\right)\frac{d}{dx}.
\end{equation}
Thus, for the one dimensional  quantum mechanical systems with both minimum length and 
maximum momentum the deformed  Hamiltonian becomes \cite{n6}-\cite{n7}
\begin{equation}
- \frac{1}{2m}\frac{d^2 \psi}{dx^2} - i \frac{\alpha}{m}  \frac{d^3\psi}{dx^3} + \frac{5 \alpha^2}{2m}
 \frac{d^4  \psi}{dx^4} +  V(x)\psi = E  \psi. 
\end{equation}

In higher dimensions, we can also 
use the standard representation for $\tilde p_i =  -i \partial_i$ and  obtain an expression for
 $p_i$, 
\begin{equation}
p_i = -i\left(1 + \alpha \sqrt{-\partial^j \partial_j} - 2\alpha^2 \partial^j 
\partial_j\right)\partial_i.
\end{equation}
Like the deformation corresponding to the existence of 
a minimum length, this deformation also influence all quantum mechanical Hamiltonians.  
Thus, in any dimension greater than one this Hamiltonian becomes 
non-local 
\begin{equation}
- \frac{1}{2m} \partial^i\partial_i  \psi - i \frac{\alpha}{m}  \sqrt{-\partial^j\partial_j}
\partial^i\partial_i \psi+ \frac{5 \alpha^2}{2m}
\partial^i\partial_i \partial^j\partial_j \psi +  V(x)\psi = E  \psi.
\end{equation}
Thus, the term proportional to $\alpha$, is a non-local term as it contains this 
non-local differential operator $\sqrt{-\partial^j\partial_j}$. So, the existence of a 
maximum momentum will being 
induce non-locality in all quantum mechanical systems. 
However, we 
can still  effectively treat this non-local correction  as a local term 
by  using  the theory of harmonic extensions of  functions from 
$R^n$ to $R^n \times (0, \infty)$ \cite{lifz1}-\cite{lifz}. 
 So, we define $\sqrt{-\partial^j\partial_j}$ on functions 
 $\psi : R^n \to R $, such that the  harmonic extension $u: R^n \times (0, \infty) \to R$ satisfies, 
$\sqrt{-\partial^j\partial_j} \psi (x) = - \partial_y u (x, y) | _{y =0}
$.
If the restriction of a harmonic function $u: R^n \times (0, \infty) \to R$ to $R^n$,
coincides with a function $\psi : R^n \to R$, then given any function $\psi $, 
it is possible to find $u$ by solving 
the Dirichlet problem defined by  $u(x, 0) = \psi (x)$ and $ \partial^2_{n+1} u (x, y) =0  $, where 
$\partial^2_{n+1}$ is the Laplacian in $R^{n+1}$, for  $x \in R^n$ and $ y \in R$.  
 Thus, for a smooth function $C^\infty_0 (R^n) $ there is a unique harmonic extension 
 $ u \in C^\infty (R^n \times (0, \infty))$. 
 If $u$ is a harmonic extension of $\psi $, then $ u_y (x, y)$
is the harmonic extension of $\sqrt{-\partial^j\partial_j} \psi  (x)$ to $R^n \times (0, \infty)$. 
So, the following result can be obtained, 
$\sqrt{-\partial^i\partial_i}^2 \psi  (x) = \partial^2_y  u(x, y) |_{y =0}
=  - \partial ^2_n u(x, y)|_{y =0} 
$. As $(\sqrt{-\partial^j\partial_j})^2 \psi (x) = - \partial^2_n \psi (x)$, so we can consistently 
define
 $  \sqrt{- \partial^i \partial_i}= \sqrt{- \partial^2_n}$. 
 
Now if we take $\psi (x) = \cos \tilde p x$, then the bounded harmonic extension 
of $\psi $ will be given by $u (x, y) = \exp -|\tilde p| y . \cos \tilde p x$, where 
$ y \in (0, \infty)$, 
$\partial^2_x u (x, y)  + \partial^2_y u(x, y) = 0
$. The action of $\sqrt{-\partial^j\partial_j} $ on $\psi  (x)$ will be given by 
$\sqrt{-\partial^j\partial_j}  \cos \tilde p x = - u_y (x, y) \left. \right|_{y =0} 
$, and so, we have $\sqrt{-\partial^j\partial_j} \cos \tilde p x =|\tilde p| \cos \tilde p x$. 
Similarly, we take $\psi  (x) = \sin \tilde p x$, and its the bounded  harmonic  as
$u (x, y) =  \exp -|\tilde p| y . \sin \tilde p x$. So, again we have $\sqrt{-\partial^j\partial_j}
\sin \tilde p x =|\tilde p| \sin (\tilde p x)$. Now we can write the action of 
$\sqrt{-\partial^j\partial_j}$ as follows, 
$\sqrt{-\partial^j\partial_j}\exp i\tilde p x =\sqrt{-\partial^i\partial_i} (\cos \tilde p x +
i \sin \tilde p x)  
$. Thus, we have $\sqrt{-\partial^j\partial_j} \exp i\tilde p x = |\tilde p| \exp i\tilde p x$. 
So, $\sqrt{-\partial^j\partial_j}$ can be effectively used as a local derivative. 
 It may be noted that if $\psi  (x) $ admits a harmonic extension $u (x, y)$, 
 such that $\sqrt{-\partial^j\partial_j} \psi  (x) = - u_y (x, 0) $,
 then  the harmonic extension of  $\partial_i \psi (x)$ will be 
$ \partial_i u (x, y)$. Furthermore, if $u \in C^2 (R \times (0, \infty))$, then  
$\sqrt{-\partial^j\partial_j}\partial_i \psi (x) = - \partial_i u_y (x, y) \left. \right|_{y =0} 
$.  So, this non-local operator commutes with a derivative,
 $\sqrt{-\partial^j\partial_j} \partial_i \psi (x) = \partial_i\sqrt{-\partial^j\partial_j} \psi (x)  $.
 \section{Quantum Field Theory}
 It may be noted the first quantized theory with minimum 
 length has also been second quantized  \cite{n9}-\cite{n}. 
In order to achieve this the deformed Heisenberg  algebra has also been extended to also include 
a minimum time. Thus, the temporal part of the first quantized theory is also deformed like its 
spatial part. So, it is a natural to complete this algebra to include a 
maximum energy. Thus, we propose the following  algebra, 
\begin{equation}
 [x^\mu, p_\nu] = i \left[  \delta_{\nu}^\mu - \alpha |p^\rho p_\rho|^{1/2} \delta_{\nu}^\mu + 
 \alpha |p^\rho p_\rho |^{-1/2} p^\mu p_\nu 
  + \alpha^2 p^2 \delta_{\nu}^\mu + 3 \alpha^2 p^\mu p_\nu\right],
\end{equation}
where $p_\mu$ stands for the full 
four momentum of a particle.
This algebra can be represented by the following deforming of the momentum operator, 
$p_\mu  = \tilde p_\mu (1 - \alpha |\tilde p^\rho \tilde  p_\rho|^{1/2}- 
2 \alpha^2\tilde p^\rho \tilde  p_\rho)$. 
We will now work in Euclidean spacetime and thus take $\tau = it$
as the time variable.
We can obtain the following expression for $p_\mu$ in coordinate representation, 
\begin{equation}
 p_\mu = -i(1 +  \alpha \sqrt{-\partial^\nu \partial_\nu} 
- 2\alpha^2 \partial^\nu \partial_\nu)\partial_\mu.
\end{equation}
Thus, we can write the deformed Klein–Gordon equation as follows, 
\begin{equation}
 \left(1 - 2\alpha \sqrt{-\partial^\nu \partial_\nu} + 5\alpha^2 \partial^\nu \partial_\nu 
\right) \partial^\mu \partial_\mu \phi + m^2 \phi =0.
\end{equation}
We can write the Lagrangian corresponding to it as follows, 
\begin{equation}
 \mathcal{L} = 
\phi \left[ \left(1 - 2\alpha \sqrt{-\partial^\nu \partial_\nu} 
+ 5\alpha^2 \partial^\nu \partial_\nu 
\right) \partial^\mu \partial_\mu  + m^2\right] \phi.
\end{equation}
In order to derive the deformed Klein–Gordon equation, from this Lagrangian, we needed 
to shift the non-local operator $\sqrt{-\partial^\nu\partial_\nu}$ from one field to the next. 
So, we  the harmonic extensions of fields $\phi_1$ and $\phi_2$ on $ C = R \times (0, \infty) $
be $u_1$ and $u_2$, respectively. 
Also let these harmonic extensions vanish for $|x| \to \infty $ and $|y| \to \infty $.
Now we  can write 
\begin{equation}
\int_C u_1(x, y) \partial_{n+1}^2 u_2 (x, y) dx dy - \int_C  u_2(x, y) \partial_{n+1}^2 u_1 (x, y) dx dy
= 0. 
\end{equation}
So, we can write 
 \begin{equation}
 \int_{R^n} \left(u_1(x, y) \frac{\partial}{\partial y } u_2 (x, y)  -   u_2(x, y) \frac{\partial}{\partial x } u_1 (x, y) \right)\left. \right|_{y =0} dx 
= 0. 
 \end{equation}
From this we get 
 \begin{equation}
 \int_{R^n}\left(\phi_1(x) \frac{\partial}{\partial y } \phi_2 (x) - 
 \phi_2(x) \frac{\partial}{\partial x } \phi_1 (x)\right)  dx 
= 0. 
 \end{equation}
 Thus, we can write 
 \begin{equation}
 \int_{R^n} \phi_1 (x) \sqrt{-\partial^\nu\partial_\nu} \phi_2 (x)  dx = 
 \int_{R^n} \phi_2 (x) \sqrt{-\partial^\nu\partial_\nu} \phi_1 (x) dx \label{l}
 \end{equation}

The Euclidean Green's function corresponding to this deformed theory can be calculated by  
nothing that the existence of a minimum length and maximum momentum, will deform the Euclidean  
 Green's function. This will occur because
 the momentum  in the exponent will get deformed to, $p_\mu =
\tilde p_\mu (1 - \alpha |\tilde p| - 2 \alpha^2 \tilde p^2)
$. 
Thus, we can write, $G(\tilde p) = [J (\tilde p )]^{-1}/ (\tilde p^2 + m^2)$, where 
$J (\tilde p ) = \det [\partial p/ \partial \tilde p ]$. 
Similarly, the propagator for any physical field  can be obtained by taking the 
Jacobian determinant of the transformation between $p$ and $\tilde p$.
It may be noted that a 
similar result has been obtained for theories with a  minimum measurable length 
\cite{n9}. 

The action of the non-local operator $\mathcal{T}_{\partial}$ on the fermionic fields 
can also be effectively defined by  using  the theory of harmonic extensions of  functions. 
Thus, we can write the Dirac equation as follows, 
\begin{equation}
 i \gamma^\mu  \left( 1 + \alpha \sqrt{-\partial^\nu\partial_\nu}  -
2 \alpha^2 \partial^\nu \partial_\nu\right) \partial_\mu \psi - m\psi = 0.
\end{equation}
Now we can obtain the following result, 
 \begin{equation}
 \int_{R^n} \bar \psi (x) \sqrt{-\partial^\nu\partial_\nu} \psi (x) dx  =  
 \int_{R^n} \sqrt{-\partial^\nu\partial_\nu} \bar \psi (x) \psi (x) dx ,
 \end{equation}
 by repeating the argument used in the derivation of a similar equation in the bosonic case. 
So, we can write the Lagrangian for the deformed Dirac's equation as, 
\begin{equation}
 \mathcal{L} = \bar \psi \left[ i \gamma^\mu 
 \left( 1 + \alpha \sqrt{-\partial^\nu \partial_\nu} - 
2 \alpha^2 \partial^\nu \partial_\nu\right)\partial_\mu  - m\right] \psi.
\end{equation}
Now  the propagator corresponding to this equation will be given by, 
$ G(\tilde p) =i [J (\tilde p )]^{-1}/ (\gamma^\mu \tilde p_\mu - m )$.
This deformation of the propagator of physical fields, 
is also a universal feature of all second quantized theories, 
whose first quantized equations have been deformed. It is important to stress that this is 
true only 
for physical fields, this is becuase in the next section, we will observe that the propagator 
for ghost fields does not get deformed in these theories. 
\section{Gauge Symmetry}
It is known that the usual undeformed Dirac's equation is invariant under a global phase 
transformation. 
This global symmetry can be promoted to a local gauge symmetry by replacing all the derivatives 
by covariant derivatives. Here the covariant derivatives are defined by introducing a gauge field 
which cancels the extra terms generated by action of the derivative on the gauge parameter. 
Furthermore, by making this gauge parameter a matrix valued function, this abelian gauge symmetry 
can be promoted to an non-abelian gauge symmetry. Finally, a 
kinetic term for any gauge theory can be obtained by taking a 
commutator of two covariant derivatives. Thus, we could proceed this way for the deformed  
Dirac's theory also. So, we could define a gauge field that would cancel the extra terms generated 
by the action of the deformed derivatives on the gauge parameter. The problem with this approach 
is that, the deformed derivatives contain a non-local part. Thus, there are effectively 
infinity many terms in this theory and it is not possible to obtain a
compact form for the non-local 
gauge transformations that a gauge field has to transform under, so that a covariant derivative can 
be constructed. 

However, if we make all the derivatives covariant, then the deformed express will 
also be covariant, 
$\mathcal{D}_\mu = ( 1 + \alpha \sqrt{-D^\nu D_\nu} - 
 2 \alpha^2 D^\nu D_\nu)  D_\mu$ and $D_\mu = 
 \partial_\mu - ig A_\mu$. In fact, we can directly incorporate non-abelian gauge symmetry at this 
 stage by deforming the derivatives as follows, $\mathcal{D}_\mu = 
 ( 1 + \alpha \sqrt{-D^\nu D_\nu} - 
 2 \alpha^2 D^\nu D_\nu) D_\mu$ and $D_\mu = 
 \partial_\mu - ig A^A_\mu T_A$, such that $[T_A, T_B] = i f^C_{AB} T_C$. 
 We can thus write a Lagrangian  which will be invariant under gauge transformations 
 as follows, 
\begin{equation}
  \mathcal{L} = \bar \psi  i \gamma^\mu \mathcal{D}_\mu  \psi - \
  \frac{1}{4} \mathcal{F}^{ \mu \nu} \mathcal{F}_{ 
 \mu\nu}, 
\end{equation}
Here the deformed field strength  $\mathcal{F}_{ 
 \mu\nu} = \mathcal{F}_{ 
 \mu\nu}^A T_A$, used
in  this theory is  obtained by taking commutator of two 
deformed covariant derivatives, 
 \begin{eqnarray}
   \mathcal{F}_{\mu\nu} &=& 
   -i[\mathcal{D}_\mu, \mathcal{D}_\nu] 
      \nonumber \\ &=&  -i [[D_\mu , D_\nu]  + \alpha [\sqrt{-D^\tau D_\tau } D_\mu , D_\nu] 
  +\alpha [D_\mu, \sqrt{-D^\tau D_\tau } D_\nu]  \nonumber \\ &&  + 
   \alpha^2 [\sqrt{-D^\tau D_\tau } D_\mu ,\sqrt{-D^\rho D_\rho } D_\nu]
   - 2 \alpha^2 [ D^\tau D_\tau D_\mu , D_\nu] \nonumber \\ && - 2\alpha^2 [D_\mu , 
   D^\tau D_\tau D_\nu]].
 \end{eqnarray}
 
 Thus, we get, $\mathcal{F}_{\mu\nu}= F_{\mu\nu} + O (\alpha)$, where 
 $F_{\mu\nu}$ is the convectional Yang-Mills field tensor. It may be noted that, 
 if under gauge transformations, $F_{\mu\nu} \to  U F_{\mu\nu} U^{-1}$, then 
 under the same gauge transformations, $ \mathcal{F}_{\mu\nu} \to  U  
 \mathcal{F}_{\mu\nu} U^{-1}$. Thus, even though $ \mathcal{F}_{\mu\nu}$ is a non-local 
 non-abelian field strength, all the derivatives in it are covariant derivatives.
 So, they all transform 
 covariant and the net result is that $\mathcal{F}_{\mu\nu}$ transforms under local 
 gauge transformations like $F_{\mu\nu}$. By the same argument,
 the whole Lagrangian for this deformed Yang-Mills theory coupled to matter fields, 
 is invariant under local 
 gauge transformations. 
It may be noted that even a deformed abelian gauge theory 
with contain infinite number of interaction 
terms due to the non-local operator.

It is not clear how we can quantize this theory, as it is not clear if the 
derivative in the ghost term will also get deformed. However, we can use the 
fact that this theory is invariant under usual gauge transformations, and thus 
the structure of the BRST and the anti-BRST transformations will not change. We can 
now use a particular non-linear gauge, where we  incorporate into
the gauge-fixing Lagrangian a quartic ghost interaction
\cite{brst}. The advantage of using this gauge is that 
 the sum of the gauge fixing term and the ghost term contains no derivatives. 
 However, the usual BRST and the usual anti-BRST transformations get deformed in this gauge, 
 \begin{eqnarray}
  s \bar c^A = b^A - \frac{1}{2}(\bar c \times c)^A, && \bar s c^A = - b^A - \frac{1}{2} 
  (\bar c \times c)^A,
  \nonumber \\ 
  s b^A = -\frac{1}{2} (c\times b)^A - \frac{1}{8} ((c\times c)\times \bar c)^A, &&
  \bar b^A = -\frac{1}{2} (\bar c\times b)^A - \frac{1}{8} ((\bar c\times \bar c)\times c)^A,
    \nonumber \\ 
    s c^A = -\frac{1}{2} (c \times c)^A,  && \bar s \bar c^A = -\frac{1}{2} (\bar c \times \bar c)^A, 
    \nonumber \\ 
    s A_\mu^A  = (D_\mu c)^A, && \bar s A_\mu^A  = (D_\mu \bar c)^A.
 \end{eqnarray}
Now we can write the sum of the gauge fixing term $  \mathcal{L}_{gf} $ and the ghost 
term $ \mathcal{L}_{gh} $ as follows \cite{brst} 
\begin{eqnarray}
 \mathcal{L}_{gf} + \mathcal{L}_{gh} &=& 
 \frac{i}{2} s\bar s (A^\mu A_\mu - i \xi \bar c c)
 \nonumber \\ &=& 
 i b\partial^\mu A_\mu + \frac{\xi}{3} b^2 + \frac{i}{2} \bar c 
 \partial^\mu D_\mu c + \frac{1}{8} \xi 
 (\bar c\times c)^2. 
\end{eqnarray}
Thus, at least in this non-linear gauge the ghost propagators do not get deformed. This is because they depend on  the 
gauge transformations, and this non-local theory is invariant under regular
local gauge transformations.

\section{Conclusion}
In this paper, we analysed a universal correction that will occur in all quantum mechanical Hamiltonians, due to the 
existence of both a minimum length and maximum momentum. These corrections occur due to the deformation of the Heisenberg algebra. 
This in turn deforms the coordinate representation of the momentum operator. This deformed momentum operator contains a non-local term, which 
reduces to a local term only in one dimension. The existence of a minimum momentum scale will induce non-locality in all quantum mechanical 
processes at high energy scales. However, this non-local term can be effectively treated as a local term by using the theory of harmonic extensions of  functions from 
$R^n$ to $R^n \times (0, \infty)$. We have 
also analysed the implications of this deformation for quantum 
field theory. Thus, we were able to construct a non-local Lagrangian for the Dirac equation. 
Then we coupled this Lagrangian to non-abelian gauge fields. We observed that
even though the field
strength corresponding  for the gauge part of the theory contains non-local terms, it transformed 
under local gauge transformations, as a regular undeformed field strength. Thus, by the same argument, 
it was argued that the whole Lagrangian for the deformed Yang-Mills coupled to matter fields, 
 is invariant under local 
 gauge transformations. The quantization of this theory was also discussed. 

It may be noted that like Yang-Mills theories, it is possible to describe 
the gravity in the framework of gauge theories. 
Thus, if we deform a covariant derivative in curved spacetime, then the 
dynamics of the gravitational field itself will change. 
In this case, the  Einstein Hilbert action will be built from a generalized Riemann
tensor which in turn will be defined though the  commutator of these
deformed covariant derivatives. This generalized Einstein Hilbert action will also 
contain non-local terms and the
variation with respect to the tetrad field yields the non-local Einstein equations. 
The first part of these equations will be the usual local  Einstein equations. 
However, it is expected that we will also obtain non-local parts proportional to $\alpha$. 
This means that the dynamics of the metric 
structure of spacetime depend  will be considerably deformed. Usually the deformation of the 
Heisenberg algebra seen as a consequence of some quantum gravitational effect. 
However, it will be also interesting to reverse this argument and analyses the effects 
a deformation of the 
Heisenberg algebra can have for the quantum theory of gravity.

\end{document}